\documentclass[]{spie}  %>>> use for US letter paper
\usepackage[]{graphicx}

\usepackage{booktabs}
\usepackage[]{graphicx}
\usepackage{wrapfig}
\usepackage[labelfont=bf]{caption}

\usepackage{amssymb}
\usepackage{bm}
\usepackage{color}
\usepackage{amsmath}
\newcommand{\degree}{^\circ}

\newcommand{\be}{\begin{eqnarray}}
\newcommand{\ee}{\end{eqnarray}}

\title{PANGU: A High Resolution Gamma-ray Space Telescope} 
\author{Xin Wu\supit{a}, Meng Su\supit{b}, Alessandro Bravar\supit{a}, Jin Chang\supit{c}, Yizhong Fan\supit{c}, Martin Pohl\supit{a} and Roland Walter\supit{d}
\skiplinehalf
\supit{a} \small\textit{Département de Physique Nucléaire et Corpusculaire (DPNC), University of Geneva, Geneva, Switzerland};\\ %\'e
\supit{b} \small\textit{Kavli Institute for Astrophysics and Space Research, Massachusetts Institute of Technology, Cambridge, U.S.A};\\
\supit{c} \small\textit{Key Laboratory of Dark Matter and Space Astronomy, Purple Mountain Observatory, Chinese Academy of Sciences, Nanjing 210008, China};\\
\supit{d} \small\textit{INTEGRAL Science Data Center (ISDC), University of Geneva, Geneva, Switzerland};\\
}
\authorinfo{E-mail to: xin.wu@cern.ch, mengsu@mit.edu}

\begin{document} 
\maketitle 

\begin{abstract}
We describe the instrument concept of a high angular resolution telescope dedicated to the sub-GeV (from $\gtrsim$10 MeV to $\gtrsim$1 GeV) gamma-ray photon detection. This mission, named {\it PANGU} ({\bf PA}ir-productio{\bf N} {\bf G}amma-ray {\bf U}nit), has been suggested as a candidate for the joint small mission between the European Space Agency (ESA) and the Chinese Academy of Science (CAS). A wide range of topics of both astronomy and fundamental physics can be attacked with {\it PANGU}, covering Galactic and extragalactic cosmic-ray physics, extreme physics of a variety of extended (e.g. supernova remnants, galaxies, galaxy clusters) and compact (e.g. black holes, pulsars, gamma-ray bursts) objects, solar and terrestrial gamma-ray phenomena, and searching for dark matter decay and/or annihilation signature etc. The unprecedented point spread function can be achieved with a pair-production telescope with a large number of thin active tracking layers to precisely reconstruct the pair-produced electron and positron tracks. Scintillating fibers or thin silicon micro-strip detectors are suitable technology for such a tracker. The energy measurement is achieved by measuring the momentum of the electrons and positrons through a magnetic field. The innovated spectrometer approach provides superior photon pointing resolution, and is particular suitable in the sub-GeV range. The level of tracking precision makes it possible to measure the polarization of gamma rays, which would open up a new frontier in gamma-ray astronomy. The frequent full-sky survey at sub-GeV with {\it PANGU}'s large field of view and significantly improved point spread function would provide crucial information to GeV-TeV astrophysics for current/future missions including Fermi, DAMPE, HERD, and CTA, and other multi-wavelength telescopes.
\end{abstract}


\section{INTRODUCTION}
\label{sec:intro}
% Earth limb as transient varibility
High energy gamma-ray photons produced in the cosmos hold the key information from the extreme of violent phenomena in the Universe to questions of fundamental physics e.g. what the dark matter particles are made from. The highly successful spaceborne gamma-ray experiments including AGILE launched in 2007\cite{AGILE:2003} and the Fermi Gamma-ray Space Telescope (Fermi) launched in 2008\cite{Atwood:2009} have brought a wealth of new information on many phenomena in gamma-ray astronomy. The Large Area Telescope on board Fermi covers a broad energy range from $\sim$100 MeV to $\sim$300 GeV. Thousands of a variety of gamma-ray emitting sources have been detected, including but not limited to blazars, pulsars, starburst galaxies, radio galaxies, nearby galaxies, globular clusters, supernova remnants, and also terrestrial gamma-ray flashes and solar flares. The gamma-ray production is often associated with physics of relativistic non-thermal process, strong gravity regimes, and plasma instabilities, which are not able to be probed in terrestrial laboratories yet. Among the detected gamma-ray sources, more than 1/3 are unidentified sources, i.e. sources without confirmed counterparts from observations of other wavelength. Such identification is crucial to reveal the nature of these sources, which requests not only increased instrumental sensitivity, but more importantly higher angular resolution to locate the sources with future gamma-ray missions. Furthermore, the gamma-ray sky is dominated by the Galactic diffuse emission produced by the interactions of high energy cosmic rays with interstellar gas and radiation fields. The spectral and spatial information of the Galactic diffuse gamma-ray emission has the potential to reveal much about the acceleration sources and propagation of cosmic rays. High spatial resolution of gamma-ray measurement is crucial to separate point source emission from the highly structured diffuse emission, thus enable reliable measurement of both the spectrum and the spatial position/distribution of the point sources. It would also allow separation of contributions from various production mechanism of diffuse gamma rays. 
%The extragalactic background, if reliably determined, can be used in cosmological and blazar studies. Studying the derived "average" spectrum of faint Galactic sources may be able to give a clue to the nature of the emitting objects.

%The Fermi and AGILE space-based telescopes are expected to continue to make significant progress for the next several years. However, there is no gamma-ray space scheduled as the successor of Fermi for the next several years from NASA or ESA. The Gamma-ray Large Area Space Telescope (GLAST) is in the conceptual development stage and is scheduled for a New Start in FY 2001 with a planned launch in 2005. The telescope will cover an energy range from 10 MeV to 300 GeV. There are currently two instrument design concepts for GLAST: SiliconGLAST and FiberGLAST. Both designs are pair production telescopes, which use a detector to track the converted pair and its associated electromagnetic shower to determine the incident direction and energy of the incoming photon. Fermi utilizes silicon strip detectors and a CsI(Tl) calorimeter.

Here we propose a high angular resolution small mission, dedicated to unprecedented precision measurement of sub-GeV ($\gtrsim$10 MeV to $\gtrsim$1 GeV) gamma-ray photons with polarimetry capabilities. This instrument concept, named {\it PANGU} ({\bf PA}ir-productio{\bf N} {\bf G}amma-ray {\bf U}nit), has been suggested as a candidate for the CAS-ESA joint small mission\footnote{\texttt{http://sci.esa.int/cosmic-vision/53072-esa-and-cas-planning-for-a-joint-mission/}}, and has the potential opening up a unique observational window of electromagnetic spectrum that has not been explored yet with great precision. A wide range of topics of both astronomy and fundamental physics can be attacked with a telescope that has an angular resolution a factor of $\sim 5$ better than the currently operating Fermi Gamma-ray Space Telescope in the sub-GeV range, covering Galactic and extra-galactic cosmic-ray physics, extreme physics of a variety of extended (e.g. supernova remnants, galaxies, galaxy clusters) and compact (e.g. black holes, pulsars, gamma-ray bursts) objects, solar and terrestrial gamma-ray phenomena, and searching for dark matter decay and/or annihilation signature etc. 

The unprecedented spatial resolution can be achieved with a pair-production telescope that, instead of the high-Z converter commonly used, relies on a large number of thin active tracking layers to increase the photon conversion probability, and to precisely track the pair-produced electron and positron tracks. Scintillating fibers or thin silicon micro-strip detectors are suitable technology for such a tracker. The energy measurement is achieved by measuring the momentum of the pair-converted electrons and positrons through a magnetic field. The innovated spectrometer approach provides superior photon pointing resolution, and is particular suitable in the sub-GeV range, where the opening angle between the electron and positron tracks is relatively large. Moreover, the level of tracking precision makes it possible to measure the {\em polarization} of gamma rays, which would open up a new frontier in gamma-ray astronomy. Here we present an overview of the current design of {\it PANGU} and the science goal for {\it PANGU} mission.

\section{Science objectives}

Not only spatial resolution is a key to separate different components from the gamma-ray sky, significant improvement in sensitivity for pair-production telescopes compared to Fermi can only be achieved through a dramatic improvement in the angular resolution, especially at $\lesssim$GeV. There is a lower limit of angular resolution of any nuclear pair-production telescope due to the undetected momentum of the recoil nucleus. This kinetic limit is a rapidly decreasing function of the energy of the photon, and amounts to $\sim 0.3\degree$ at 100 MeV and $\sim 0.02\degree$ at 1~GeV. 

{\it PANGU} will provide the first high resolution full-sky map in the energy range from $\gtrsim$10 MeV to $\gtrsim$1 GeV. {\it PANGU} will survey the full sky with a size of 68$\%$ containment of the point spread function (PSF) $\lesssim$1$\degree$ at $\sim 100$ MeV, which is $\gtrsim$5 times better than Fermi-LAT (the expected best available maps in this energy range by $\sim$2020) and only a factor of $\sim 3$ times of the kinetic limit. The high quality gamma-ray maps will significantly improve the identification and separation of point sources from extended and complicated diffuse gamma-ray background. It also opens up the discovery window for new phenomena in the soft gamma-ray region.

The science goals of {\it PANGU} cover a wide range of topics of Galactic and extragalactic astronomy, as well as key issues of fundamental physics. The unique science goals can only be achieved with either {\it PANGU}'s high angular resolution at sub-GeV or {\it PANGU}'s potential of polarization detection. In some cases, both of the two unique capabilities of {\it PANGU} are required. We list the major science goals of {\it PANGU}, including but not limited to:
%% can be classified into two categries, one catagry relies on {\it PANGU}'s high angular resolution at sub-GeV, one class relies on {\it PANGU}'s polarization sensitivity. 

\begin{itemize}
 \item Galactic and extragalactic cosmic rays (origin sites including supernova remnants and superbubbles, and the acceleration/propagation mechanisms)
 \item Galactic and isotropic diffuse gamma-ray emission, including the Fermi bubbles in the Milky Way\cite{FermiBubble}; separation of various diffuse gamma-ray components
 \item Search for dark matter decay/annihilation produced gamma-ray emission from the Milky Way dark matter halo, e.g. the claimed diffuse GeV excess in the inner Galaxy\cite{Daylan:2014}, or extra-galactic dark matter structures
 \item Detect and determine soft gamma-ray spectrum of compact sources, including gamma-ray bursts (GRBs), millisecond pulsars (pulsar wind nebulae), blazars, and microquasars
 \item Resolving the Galactic plane, particularly the Galactic Center region in GeV, separating point sources (Sgr A$\star$ and pulsars) and SNRs from the diffuse emission
 \item Extreme physics of black holes at different mass scales
 \item Solar and terrestrial high energy phenomena, e.g. flares and cosmic ray acceleration in solar winds
 \item First detection of polarization of gamma rays from astronomical objects
 \item Origin of ultra-high energy cosmic rays
 \item Fundamental physics, e.g. baryon asymmetry in early universe, birefringence effect arising in quantum gravity
 \item Full-sky monitoring of a variety of soft gamma-ray transients (Crab Nebula flares, blazar flares, solar flares) and significantly improved localization of gamma-ray counterparts to potential gravitational wave transients detected by the Advanced LIGO/Virgo Observatories
\end{itemize}

Sub-GeV is well-known as the crucial energy region to distinguish the hadronic from leptonic scenario of gamma-ray production in e.g. supernova remnants\cite{Ackermann:2013Sci}. With the significantly improved spatial resolution, {\it PANGU} is able to identify the characteristic $\pi^0$-decay gamma rays from that produced via bremsstrahlung and inverse Compton scattering of electrons, and avoid significant contamination from bright Galactic diffuse emission. The $\gtrsim$10 MeV low energy coverage has the potential to detect the $\pi^0$ bump in the gamma-ray spectrum, which is a key to identify the $\pi^0$ gamma rays. 

Source identification is important to enable precise measurement of the Galactic and extra-galactic diffuse gamma-ray background, which is crucial not only for studying its origin from cosmic-ray interaction with ISM and radiation field, but also for searching induced gamma-ray emission from decay and/or annihilation of dark matter particles. Recent evidence of a $\sim$GeV excess from the Galactic center region has been claimed as a potential signal from dark matter annihilation\cite{Hooper:2013}. The revealed spectrum and spatial distribution is consistent with what a few tens of GeV Weakly Interacting Massive Particle (WIMP) annihilate to standard model particles in the inner Galactic halo. Despite the claimed quite significant detection, systematic uncertainties due to contamination from diffuse emission along line of sight and unresolved point sources could be significant and hard to estimate, because of the relatively large point spread function of Fermi-LAT at $\lesssim$1 GeV, even with improved PSF by selecting well-reconstructed front-converting events\cite{Daylan:2014}. Sub-GeV is crucial to distinguish both the spectrum and the overall spatial distribution of this potential dark matter signal from millisecond pulsar models. With significantly improved PSF {\it PANGU} will resolve and separate potential gamma-ray sources in this region, thus reveal the nature of this excess by enabling more reliable gamma-ray analysis towards inner Galaxy. Furthermore, indirect dark matter search in sub-GeV gamma ray is also complementary to direct search for low mass dark matter candidates, which is not yet well constrained by other means.

As a result of the significant improvement in angular resolution, many new soft gamma-ray sources can be detected by {\it PANGU} with $< 1'$ localization precision. It's also highly complementary to gamma-ray sources detected by Fermi-LAT at $\gtrsim$GeV. High spatial resolution is crucial to reduce the contamination from diffuse emission to source separation, especially sources close to the Galactic plane at low latitude. Sub-GeV spectrum holds the key to distinguish different particle acceleration models of millisecond pulsars and detect gamma-ray emission from galaxy clusters. Other phenomena, such as blazars, gamma-ray burst, Pevatron variability in Crab nebula, gamma-ray absorption in AGN and gamma-ray loud binaries can be studied with such an instrument. Sub-GeV gamma rays are also produced in our solar system. {\it PANGU} will resolve solar flares at sub-GeV, and allow us to study terrestrial gamma-ray flashes discovered by Fermi.

Polarimetry at sub-GeV region has never been done before, {\it PANGU} will open up a whole new window of discovery. Gamma-ray polarization can distinguish between emission processes such as synchrotron radiation and other gamma-ray production mechanisms. For example polarization in the sub-GeV band offers an unique window to understand the origin of extragalactic Ultra-High Energy Cosmic Rays\cite{ZhangBottcher:2013}. High polarization is a signature of hadronic acceleration while leptonic inverse Compton emission remains unpolarized. {\it PANGU} aims to measure the sub-GeV polarization during blazar flares, which has the potential to resolve a long-standing question in both astrophysics and fundamental physics. Polarization measurements with {\it PANGU} can further being used to probe the geometry of gamma-ray emitting regions in pulsars, pulsar wind nebula, GRBs etc.  

{\it PANGU} can explore fundamental questions in physics with it's low energy extension to $\sim 10$ MeV and high angular resolution. The cosmic gamma-ray background between a few MeV and a few 100 MeV is very poorly known\cite{Inoue:2013}. Different source categories are expected to contribute: Supernovae, Flat Spectrum Radio Quasars (FSRQs), blazars, star forming galaxies, cluster of galaxies. The situation is very unclear, also because the emission of these sources is largely unknown in that spectral band. Observations of the sky over large angular scales with good angular resolutions are key to decipherate the origin of the background and understand if any part of it remains unexplained (in particular the bump between 10 and 100 MeV). Annihilation of anti-baryons in the early Universe should have left a broad gamma-ray emission component redshifted to the 10-100 MeV band\cite{Cohen:1998}. The spectral shape and the anisotropies\cite{Gao:1990} of that emission depend on the baryon asymmetry\cite{Stecker:2002}, particle transport through magnetic field, redshift and scales at which the annihilation took place. The gamma-ray spectrum can thus give important constrains on the strength of the baryon asymmetry on scales beyond clusters of galaxies\cite{vonBallmoos:2014}.

{\it PANGU} can also explore fundamental questions in theoretical physics with polarization sensitivity. Theories of quantum gravity predict a potential signal from vacuum birefringence, i.e. photons with different polarizations travel at slightly different velocities. Such a process from distant astronomical sources with the accumulated effect might destroy the inherent polarization of the sources. The capability of polarization detection of gamma-ray photons from distant sources e.g. GRBs can provide sensitive constraints the possible existence of violations of relativity\cite{Fan:2007}. 

Finally, the sub-GeV full sky survey by {\it PANGU} provides crucial complementary information to GeV to hundreds of TeV observations by Fermi, DAMPE\footnote{\texttt{http://dpnc.unige.ch/dampe/index.html}}, HERD\footnote{\texttt{http://herd.ihep.ac.cn/}}, Gamma-400\cite{Galper:2013}, HAWC\cite{HAWC}, Cherenkov Telescope Array\cite{Actis:2011}, and LHAASO\cite{LHAASO} etc. We note that none of these instruments has high resolution capability at $\lesssim$GeV. The sub-GeV observations by {\it PANGU} will play significant and unique role in multi-wavelength studies across the electromagnetic spectrum with other space and ground telescopes involving radio, optical, IR, X-ray, and neutrino detectors. During the operation of {\it PANGU} after $\sim$2020, next generation of telescopes including SKA, LSST, Euclid, eROSITA and PINGU etc. will also be operating. Moreover, full-sky monitoring capability of {\it PANGU} provides potential ability of electromagnetic counterpart identification of gravitational wave detection by advanced LIGO/Virgo. 
%\cite{LIGOVIRGO}

%% \begin{figure}[!h]
%% \begin{center}
%% \includegraphics[scale=1.1]{fig1.jpg}
%% \caption{\footnotesize Sketch showing the concept of modularity of a Laue lens. The lens is made of crystals placed in concentric 
%% rings, from the inner part (diffracting the higher energies) to the outer radius (lower energies). In the LAUE project, 
%% a petal as part of a whole lens will be built and test (blue part).}
%% \label{fig:completelens} 
%% \end{center}
%% \end{figure} 

\begin{figure}[!h]
\begin{center}
\includegraphics[scale=0.3]{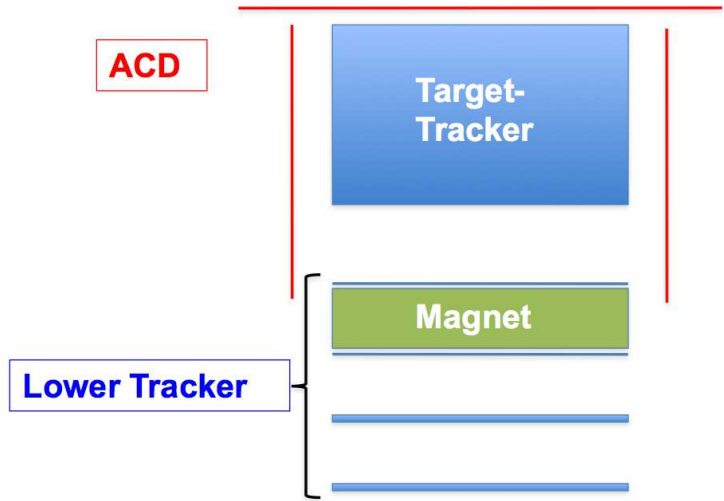}
\caption{\footnotesize The {\it PANGU} instrument is composed of three main parts: the target-tracker system, the magnetic spectrometer, and the anti-coincidence detector. }
\label{fig:layout} 
\end{center}
\end{figure}

\section{Detection principle}

Below $\sim$10 MeV, Compton scattering dominates the photon-nucleus interaction, and the gamma-ray detection relies on the multiple Compton scattering of incoming gamma-ray photons with detector materials. However, at $\gtrsim$10 MeV, pair production starts to dominate the photon-nucleus interaction over Compton scattering for low Z material. Small cross section of pair production requires more encounter material (higher radiation length) for good acceptance, while material is also the limiting factor of angular resolution because of important multiple scattering. These are two competing effects: more material is required to get photon converted, but once the electron/positron pair being produced, less intervening material before the pair tracks being measured allowing better angular reconstruction. To achieve $<1 \degree$ PSF at 100 MeV, passive material should be minimized and active detector should be thin or has low density. To increase effective area (by increasing radiation length), many layers or a large volume is required.

Previous concepts for high resolution pair telescope for gamma-ray detection include:

\begin{itemize}
\item Low density gas Time Projection Chamber (TPC): e.g. HARPO\cite{HARPO:2014} and AdEPT (5-200 MeV)\cite{AdEPT}. It potentially provides very good angular resolution, but requires large pressure vessels.
\item All silicon detectors. There are several proposals with designs optimized as Compton telescopes (with calorimeters), including MEGA/GRM\cite{MEGA} (double-sided silicon stripped detectors (SSDs), each silicon layer has a thickness of 500 $\mu$m with 5 mm distance between layers), CAPSiTT\cite{CAPSiTT} (double-sided SSDs, each silicon layer has a thickness of 2 mm with 1 cm distance between layers), TIGRE\cite{TIGRE} (double-sided SSDs, each silicon layer has a thickness of 300 $\mu$m with 1.52 cm distance between layers), and Gamma-Light\cite{Gammalight} (single-sided SSDs, each silicon layer has a thickness of 400 $\mu$m with 1 cm distance between layers).
\item Scintillating fiber, previous concepts with converters include SIFTER\cite{SIFTER} and FiberGLAST\cite{FiberGLAST}.
\end{itemize}

{\it PANGU} is a dedicated pair telescope with thin and low-Z tracker material. The payload is composed of three main parts: the target-tracker system, the magnetic spectrometer (or calorimeter), and the anti-coincidence detector (see Figure \ref{fig:layout}). 

For {\it PANGU}, both silicon and fibers trackers are viable technologies. Challenges are mainly from engineering: how to optimally use the limited weight (ultra-light module) and power budgets (low power ASICs). Silicon has been successfully used in previous space missions e.g. Fermi-LAT, AGILE, PAMELA, and AMS-02 etc. On the other hand, the cost of fiber tracker is lower, and fiber is less fragile with more flexible geometry. But the technologies of scintillating photon detector (SiPM) and readout ASICs are relatively new. There have been recent developments in high energy physics experiments, e.g. LHCb\cite{LHCb} and Mu3e etc, also on balloon experiments e.g. prototype Proton Electron Radiation Detector Aix- la-Chapelled (PERDai of Positron Electron Balloon Spectrometer (PEBS)~\footnote{\texttt{http://www1b.physik.rwth-aachen.de/~schael/PEBS.html}}, which is a balloon borne spectrometer for measuring positron/electron spectra at the TeV energy scale. We note that the position resolution $\sim$70 $\mu$m can be achieved with fiber trackers.

%Scintillating fibers were first used in space on the The Cosmic Ray Isotope Spectrometer (CRIS) instrument onboard the Advanced Composition Explorer (ACE) spacecraft launched in 1997~\cite{CRIS}. In CRIS, the trajectory of the particles are determined by the Scintillating Optical Fiber Trajectory (SOFT) system with 3 x- and y-layers of SOFT fibers for trajectory redundancy. FiberGLAST also utilizes scintillating fibers for both a tracker and a calorimeter. 
For current design, the tracker of {\it PANGU} is composed of 50 detector modules. Each module includes two orthogonal layers of scintillating fibers with an active area of 50$\times$50 cm$^2$ used to detect the passage of ionizing particles. Plastic honeycomb or foam material is used as a substrate and for the mechanical support of the module. Note that we do not add even thin conversion layers of e.g. tungsten or tantalum foil to further stimulate pair production efficiency. The maximal minimization of multiple scattering improves the angular resolution significantly.

This {\it PANGU}-Fi design is a new all-fiber tracker concept. It makes use of plastic scintillating fibers that emit characteristic light pulses upon encountering charge particles and silicon photomultipliers (SiPM) for readout. An incident gamma-ray photon will interact with the scintillating fiber and induce pair production. The resulting electron-positron pair will be detected by the tracker system itself which will measure the x-y position of the tracks at each detector layer. %% As the pair proceeds through the the 50 layers, the electromagnetic shower will be detected and can be reconstructed in three dimensions. This would allow for better angular resolution and less coulomb scattering for photons in the low-energy regime (<~100MeV)

For energy measurement, the standard way is to use a calorimeter below the tracker. For example the AGILE CsI mini-calorimeter of has a size of 37$\times$37$\times$3 cm$^3$ (corresponding to 1.5$X_0$ in depth) and weighs $\sim$30 kg\cite{AGILECAL}. However, it has a limited energy resolution of $\sim$70$\%$ at 100 MeV because of the energy leakage. With a payload limited to $<$100 kg, a calorimeter seems not an optimal option for {\it PANGU}..

The new approach we propose is to use a magnetic spectrometer built with permanent magnet below the tracker, instead of a calorimeter. By measuring the deflection angles of the electron and positron tracks due to magnetic field the moment of the particle can be incurred.  The magnet and the tracker-target system are decoupled in this configuration so they can be smaller and independently optimized. 

The four possible designs we have considered are summarised as the following:

\begin{itemize}
\item SiTCal: Silicon tracker plus crystal calorimeter. This is a classical approach, but too heavy for a small mission.
\item FiTCal: Fiber tracker plus crystal calorimeter. This is a low cost approach, but also too heavy.   
\item SiTMag ({\it PANGU}-Si): Silicon tracker plus magnetic spectrometer. It's possible to fit into a small mission, with higher technology readiness, but with higher cost.
\item FiTMag ({\it PANGU}-Fi): Fiber tracker plus magnetic spectrometer. It's a relatively new technology with lower cost, could be pioneer for future applications.  
\end{itemize}

\section{Boundary Conditions from the Joint Mission Requirements}

The European Space Agency's (ESA) Directorate of Science and Robotic Exploration (ESA-SRE) and the Chinese Academy of Science (CAS) have agreed to explore the possibility of identifying a scientific space mission which could be jointly implemented by ESA-SRE and the Chinese National Space Science Centre (NSSC) under the CAS\footnote{\texttt{http://sci.esa.int/cosmic-vision/53072-esa-and-cas-planning-for-a-joint-mission/}}. The current design of {\it PANGU} is optimized under boundary conditions of the joint mission. 

%As it is usual for ESA science missions, the ESA Member States are assumed to provide (partly or fully) the European contribution to the payload elements
Preliminary technical guidelines are:
\begin{itemize}
\item Payload mass $\lesssim$60 kg, Spacecraft launch mass $\lesssim$300 kg
\item Payload power $\lesssim$65 W average (typical)
\item Operational lifetime of satellite 2 to 3 years.
\item Technology readiness requirements: Technology readiness level (TRL) 6/7 preferable, TRL5 acceptable\footnote{TRL5: Component and/or breadboard validation in relevant environment}, by the time of the call
\item From one to several new technologies might be involved in the demonstration.
\end{itemize}

It has been suggested that the basic technological elements must be integrated with reasonably realistic supporting elements so that the total applications (component-level, sub-system level, or system-level) can be tested in a simulated or somewhat realistic environment. For {\it PANGU}, silicon slip detector has been demonstrated in previous gamma-ray space missions, and is in the TRL6/7 category, while scintillating fiber is in category TRL5. 

{\it PANGU} as a proposed candidate mission is able to comply with the pre-defined programmatic constraints by ESA and CAS, thus enable a feasible, cost-constrained mission with the expected launch no later than 2021.

\begin{figure}[!h]
\begin{center}
\includegraphics[scale=0.6]{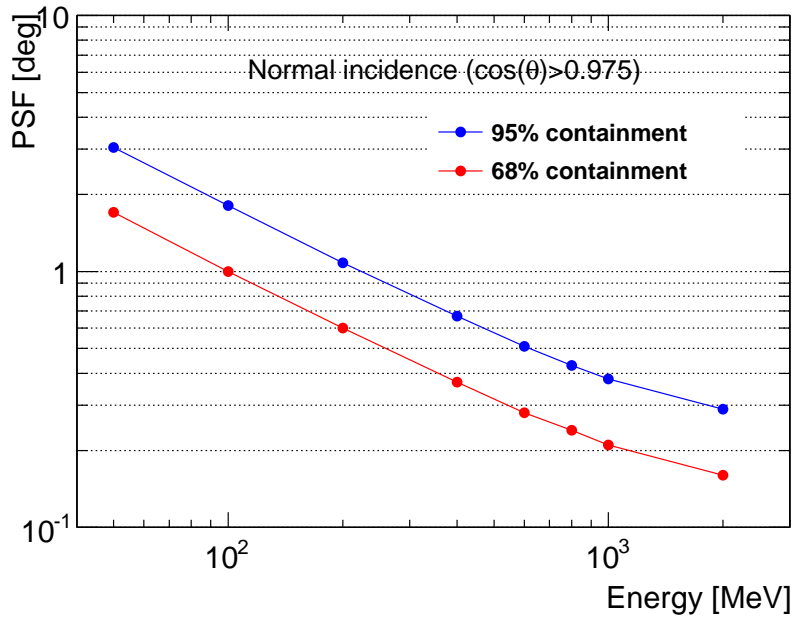}
\caption{\footnotesize The point spread function (PSF) of the current design of {\it PANGU}. The red and blue curve shows the equivalent 1$\sigma$ (68$\%$) and 2$\sigma$ (95$\%$) of the PSF respectively. }
\label{fig:psf} 
\end{center}
\end{figure} 

\begin{figure}[!h]
\begin{center}
\includegraphics[scale=0.5]{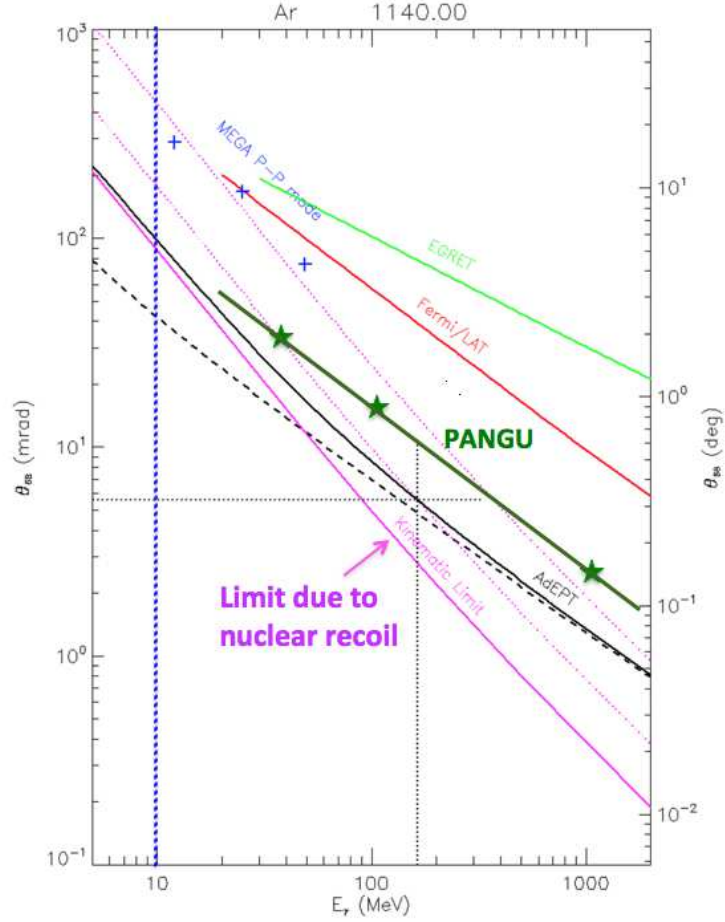}
\caption{\footnotesize (Adapted from reference \cite{AdEPT}) The expected angular resolution of the {\it PANGU} instrument concept (solid dark green line, see also \ref{fig:psf}) is shown as a function of the gamma ray energy. We show the kinematic limit for reference, along with twice and five times the limit, in terms of the 68$\%$ containment radius, for nuclear pair production\cite{Bernard:2013a,AdEPT} (solid and dotted magenta lines respectively). For energy $\lesssim$ GeV, {\it PANGU} telescope will achieve angular resolution within a factor five of the kinematic limit, and towards only a factor of two higher than the kinetic limit at tens of MeV. The MEGA\cite{MEGA} measured pair production angular resolution (blue crosses), EGRET\cite{1993ApJS...86..629T} calibrated angular resolution (green line), AdEPT\cite{AdEPT} (black solid line), and Fermi-LAT\cite{Atwood:2009} front on-orbit angular resolution (red line) are shown for comparison.}
\label{fig:psfcompare} 
\end{center}
\end{figure} 

\section{Suggested Payload}

\begin{figure}[!h]
\begin{center}
\includegraphics[scale=0.6]{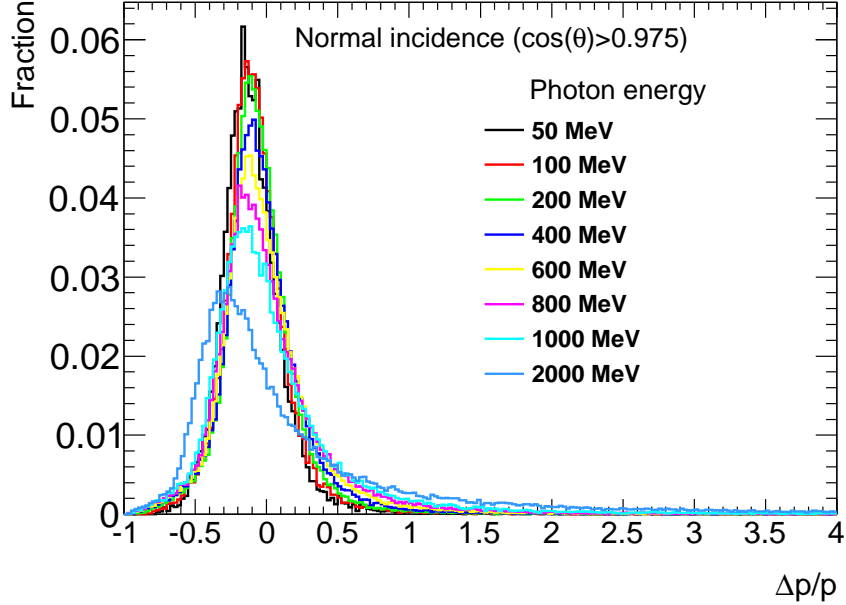}
\caption{\footnotesize Energy resolution of {\it PANGU} from 50 MeV to 2 GeV at normal incidence angle, B $=$ 0.1T. For E $<$ 1 GeV, the energy resolution is better than 20-30$\%$. }
\label{fig:deltap} 
\end{center}
\end{figure}

\begin{figure}[!h]
\begin{center}
\includegraphics[scale=0.6]{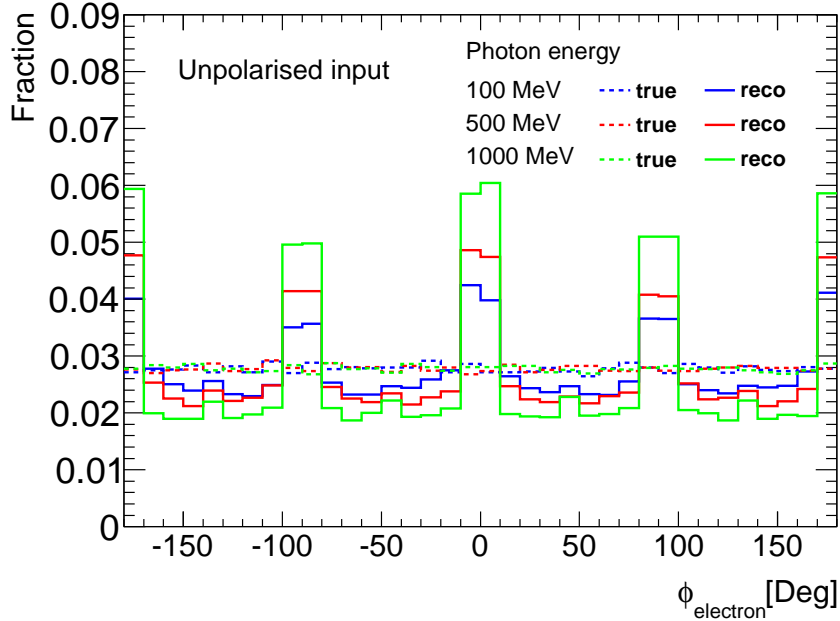}
\caption{\footnotesize The azimuthal angle $\Phi$ distribution for unpolarized input gamma-ray photons. The colored dashed lines show the input uniform distribution of $\Phi$. The colored solid lines show the corresponding reconstructed $\Phi$ distribution. We show examples at energy 100 MeV, 500 MeV, and 1 GeV. }
\label{fig:phirot} 
\end{center}
\end{figure}

\begin{figure}[!h]
\begin{center}
\includegraphics[scale=0.4]{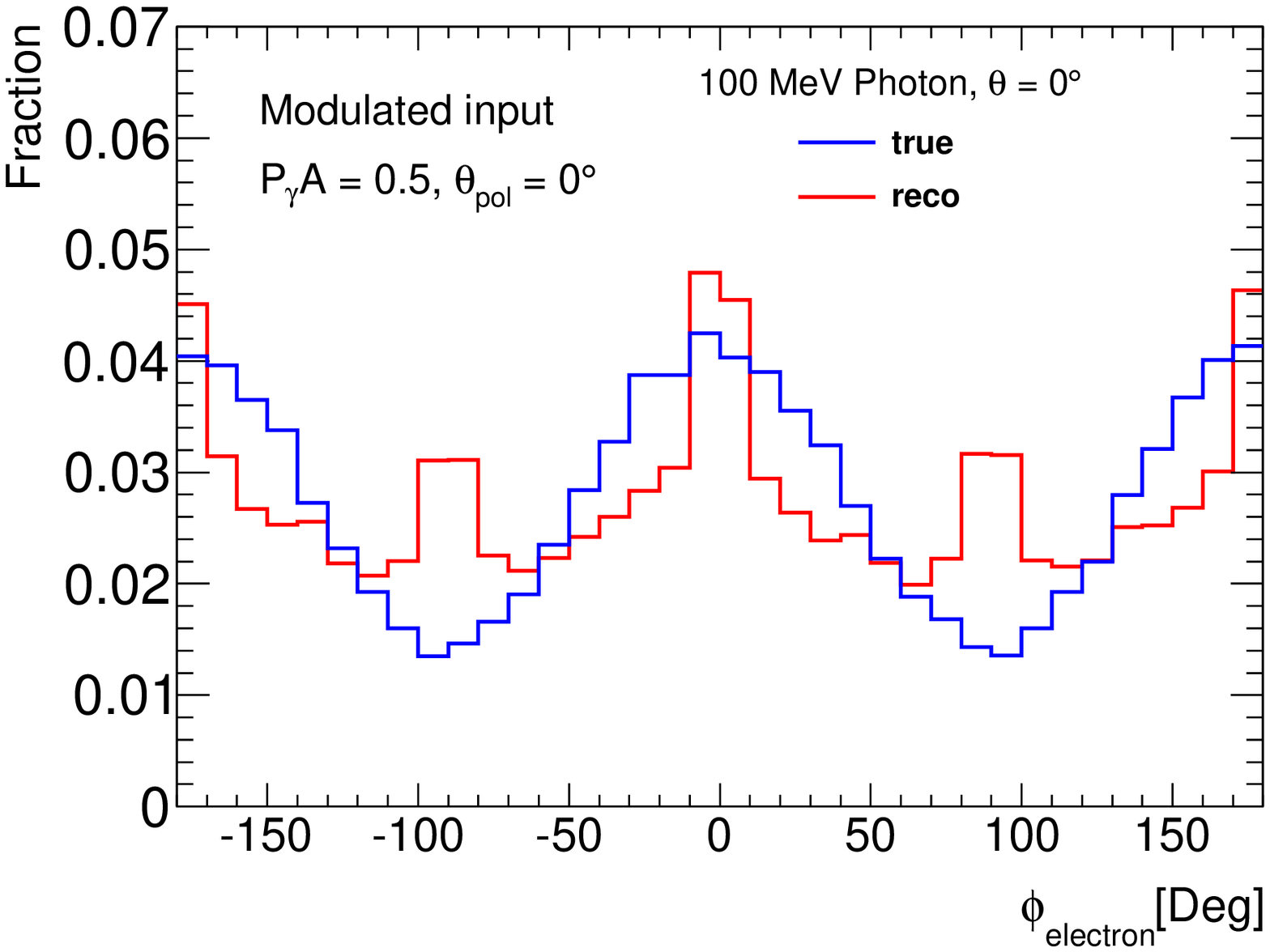}
\includegraphics[scale=0.4]{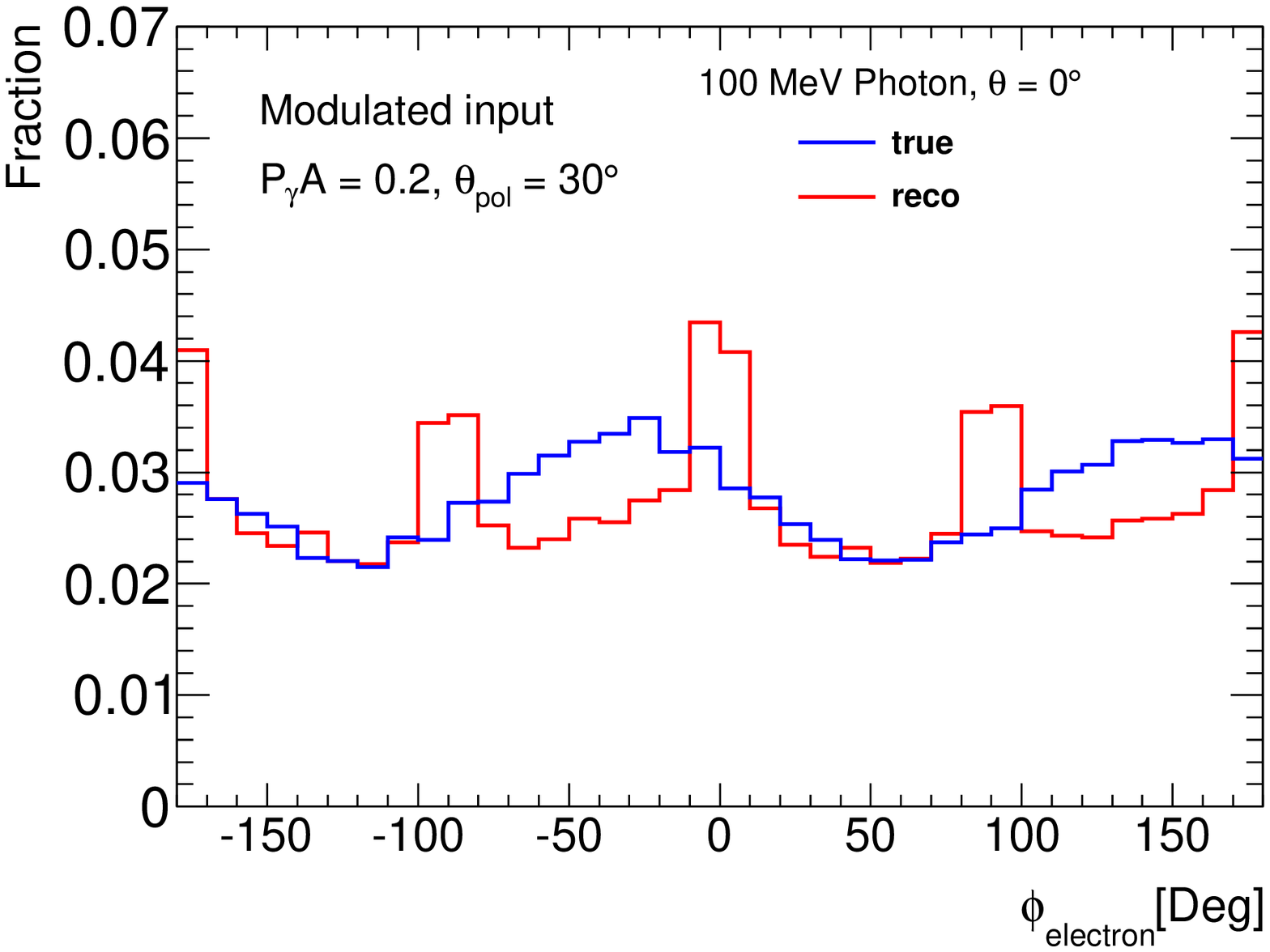}
\includegraphics[scale=0.4]{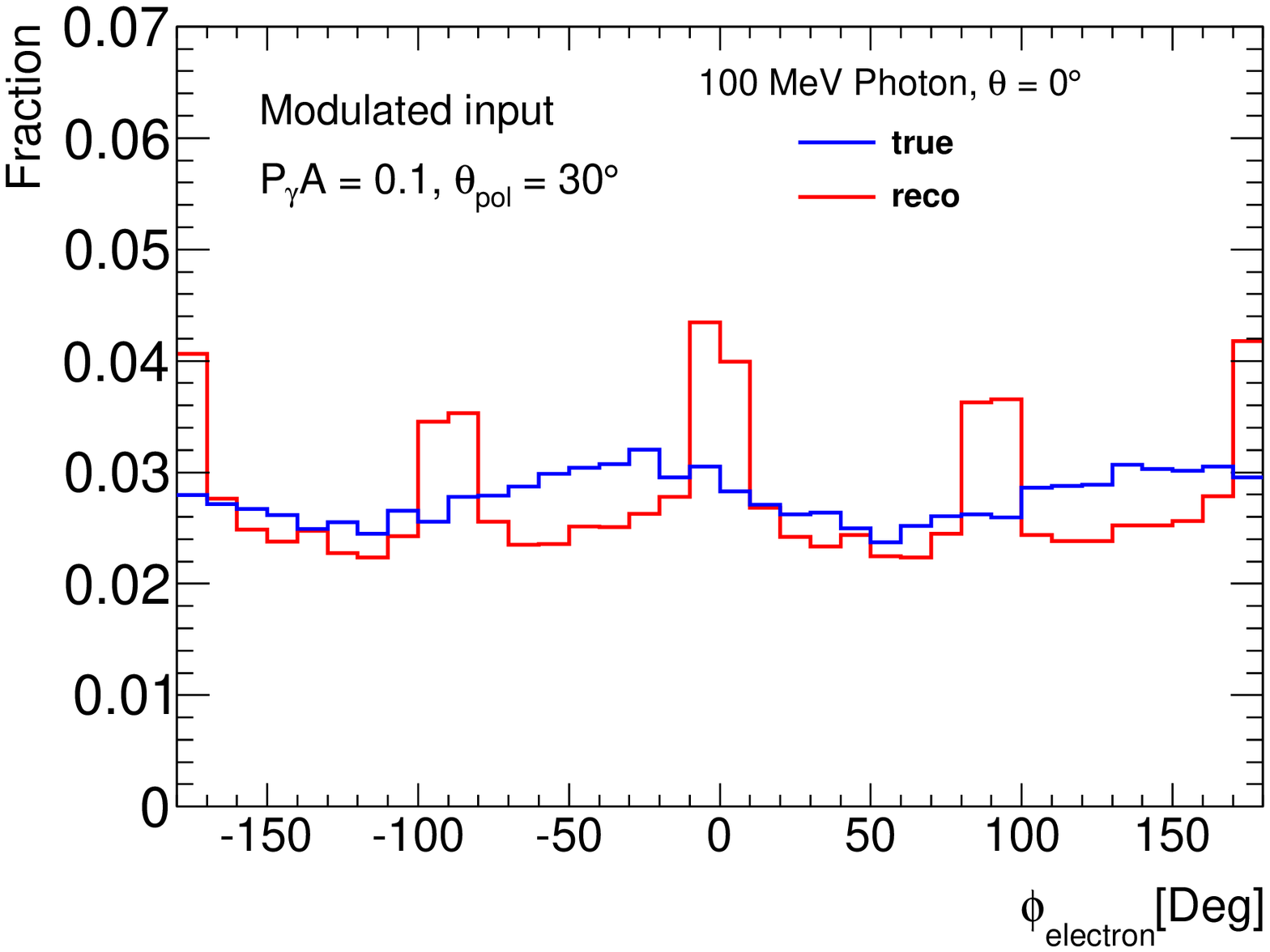}
\caption{\footnotesize  The phi distribution with different degrees of polarization: 50$\%$ (upper left), 20$\%$ (upper right), 10$\%$ (lower)}
\label{fig:pola3} 
\end{center}
\end{figure}

As introduced in Section 3, the suggested payload is a spectrometer with a fully active tracker. {\it PANGU} has three sub-systems: the target-tracker system, the magnet and lower tracker system, and the anti-coincidence system, as shown in Figure \ref{fig:layout}. The main performance specification is PSF $\lesssim 1\degree$ and energy resolution $\lesssim$30$\%$ at 100 MeV. We show the GEANT4 simulation results of {\it PANGU}'s PSF in Figure \ref{fig:psf}, for events where both electron and positron tracks are reconstructed by the spectrometer, and we compare {\it PANGU} with other missions in Figure \ref{fig:psfcompare}. The expected energy resolution is shown in Figure \ref{fig:deltap}. For a detector layout in which the tracker has a size of 50$\times$50$\times$30 cm$^3$, the inner radius of the magnet at 25 cm, magnet thickness at 5 cm, lower tracker thickness at 5 cm, the corresponding geometry factor is around 2000 cm$^2$sr. For the design of 50 double layers of fibers, the vertical thickness is $\sim$0.15X$_0$, which we use as the average thickness to calculate the conversion probability ($\sim$0.11), resulting in an effective geometry factor (GF) of $\sim$200 cm$^2$sr. The e/p the performance is expected to be similar to Fermi or AGILE at $\sim 10^5$. The polarization sensitivity is limited by the intrinsic asymmetry of the detector. Given the current design, we show in Figure \ref{fig:phirot} the azimuthal angle $\Phi$ distribution for unpolarized input gamma-ray photons. In Figure \ref{fig:pola3}, we show the possibility to detect polarization with {\it PANGU} assuming the gamma-ray source with different polarization fraction. There are four peaks of $\Phi$ corresponding to the x and y direction of the detector strips where the reconstruction of the azimuthal angles is very hard, but if it is possible to rotate the detector, e.g. by 45$\degree$, we could shift the phase of the peaks which allow us to measure all the $\Phi$ region. We note that reliable simulation code for polarized pair production is not well studied, this preliminary analysis requires further detailed study (see reference\cite{Bernard:2013b} for a recent discussion on polarimetry of the pair production process).

%The key performance requirement of {\it PANGU} is the $\lesssim$1$\degree$ level angular resolution at 100 MeV with polarization detection capability. At this level the usual pair telescope concept that uses high-Z converter is no longer adequate. Instead, a tracker consists of layers of low-Z tracking detector and minimum inactive material should be used. To arrive at a efficient conversion from photons to pairs, the total materials in the tracker, i.e. the number of tracking layers should be in the order of 10$^2$. The large number of layers drives the total number of readout channels to several hundred thousands, since each tracing layer need to be finely segmented to provide the needed position resolution of better than 100 $\mu$m. However since typically there are only 2 tracks in the event, it is possible to reduce the number of channels in each layer by sharing readout electronics for channels in different regions together, and then to rely on different connection schemes of different layers to resolve the ambiguities. The first tracking layer can serve as anti-coincidence detector (ACD). A layer of 1 cm thick plastic scintillator ACD will be used to cover the five sides of the detector.

\subsection{Tracker system}

There are two options: the baseline option is a scintillating fiber tracker (FiT), and an alternative option is a silicon tracker (SiT) using thin silicon strip detector (SSD). For an electron or positron of 100 MeV, a 1 mm scintillating fiber tracking layer generates $\sim$0.30$\degree$ {\it rms} due to multiple scattering, while a SSD of 320 $\mu$m thick generates $\sim$0.36$\degree$ {\it rms}. Both detectors can achieve the position resolution of $<$100 $\mu$m. 
%For a tracker of 0.5$X_0$ thick, the number of layers would be 200 for FiT and 150 for SiT. The total number of channels will be 120k for SiT and 150k for FiT, taking into account a reduction factor of 2 using readout sharing scheme. Current commercially available SSD readout ASICs can already reach 0.3 mW/channel, with some optimization it is reasonable to expect 0.2 mW/channel. Similar performance can be expected for FiT readout ASIC with some development effort. Thus the total power consumption of the payload can be fit into the pre-defined mission budget of 50 W. The weight of the tracker material (excluding support structure and electronics) would be around 20 kg for FiT and 11 kg for SiT.

Here we describe our current design of a possible layout of the tracker-target system. It has x-y double layers with 6 mm inter-distance, in total 50 double layers. The requirement is that each layer measures X and Y to $\sim$70 $\mu$m, total material every 6 mm is $\sim$0.3$\%$ $X_0$.  For the tracking layer with $\sim$0.3$\%$ $X_0$ in total, in case of SciFi we need two layers of 0.65 mm each (Polystyrene equivalent), each layer formed by a stack of 3 layers of size 250 $\mu$m fibers with the readout by Silicon photomultiplier (SiPM); in case of Silicon tracker, the silicon layer consists two single sided SSD of 150 $\mu$m each. We note that the total tracker active material is very small, for fibers is $\sim$15 kg (polystyrene density $\sim0.9$ g/cm$^3$), and for silicon tracker is $\sim$9 kg (silicon density ~2.33 g/cm$^3$). Both of the design needs support substrate, which is probably more for silicon for biasing, bonding, as it's more fragile. For {\it PANGU}, we aim for a total of $\sim$50 kg for fiber/silicon, support structure, electronics, ACD and possible shielding.

\subsection{Permanent Magnet Spectrometer}
%, because the conversion can be cleanly identified as two arcs that intercept tangentially, and the tangent at the intercepting point is the photon direction.
For the energy measurement, we propose an innovated concept: a spectrometer based on permanent magnets to measure the momentum of the electron and positron, the sum of the two gives the energy of the incoming photon, since the nuclear recoil energy is negligible. Measuring the electron and position tracks in a magnetic field will also improve the identification of a photon conversion event, and improve the precision of the photon angular resolution. Without energy measurements of each track but relying only on the identification of the higher energy track of the two (eg. by straightness or track length), an extra error of $\sim$0.65$\degree$ on photon PSF will be introduced for 100 MeV gamma-rays.  Since the goal of {\it PANGU} is to measure tracks $\lesssim$1 GeV, it is not necessary to require a strong magnetic field. Given the good position resolution of the {\it PANGU} tracker, even a magnetic field of 0.05 T is sufficient to achieve a momentum resolution of $\lesssim$50$\%$ at 1 GeV. With existing magnet technology it is possible to produce the required dipole field with $\lesssim$30 kg of weight, by arranging blocks of NdFeB permanent magnets according to the Halbach scheme with a cylindrical configuration.

Momentum measurement through bending angle can be obtained by:  $\theta$ = eLB/p = 0.3L$\times$B/p [mm T MeV$^{-1}$] = 3/p radian (p in MeV), where L is the traversed length, B is the strength of the magnetic field, p is the momentum of the charged particle. For 1 GeV electron/positron, the bending angle $\theta$ is 3 mrad (0.17$\degree$), and for 100 MeV it is 30 mrad (1.7$\degree$). The uncertainty of the reconstructed momentum can be estimated by: $\Delta$p/p = p/(0.3 L$\times$B) $\Delta \theta$ = (p/3) $\Delta \theta$ (p in MeV). As in the case of target-tracker, $\Delta \theta$ is dominated by tracking resolution ($\sigma_x$/d ) at high energy, and by multiple scattering at low energy. $\Delta$p/p to the level of $\sim$30$\%$-50$\%$ is feasible with B = 0.1 T and L = 10 cm, $\sigma_x$=70 $\mu$m, d = 10 cm for p = 100-1000 MeV electron/positrons, as shown in Figure \ref{fig:deltap}.  To increase the field of view (FOV), we also consider B = 0.2 T and L = 5 cm, which reduces the thickness of the detector, with the cost of doubling the weight of the magnet. However the field uniformity of such a shallow magnet needs to be studied. We could further increase the FOV by using 150 $\mu$m thick single sided SSD for lower tracker along with 100 $\mu$m readout pitch ($\sigma_x$=30 $\mu$m), and reduce the distance between the tracking layers of the spectrometer from 10 cm to 5 cm. Preliminary investigation shows that it is possible to build a dipole magnet system of 0.2 T weighting $\lesssim$30kg. The current design of the magnet has a outer/inner radius of $r_2$ = 29 cm, $r_1$ = 25 cm, with a height of 5 cm. The magnetic field direction is in +y direction. The lower tracker has one X-layer above, one X-layer, and two X-Y layers below, $\sim$10 cm between layers. The spectrometer tracking layer will be made from high precision silicon strip detectors. 

\section{Power Consumption}
The total number of readout channels of 50 double-layers in the target plus 6 layers in the lower tracker system below the magneto-spectrometer, with 250 $\mu$m readout pitch, is $\sim$2$\times$10$^5$ (1000 per single layer). The main power consumption comes from the SiPM readout ASIC. With some optimisation effort, it is possible to achieve $\sim$0.2 mW/channel, similar to the current performance of ASICs for Si strip detectors, resulting in a total ASIC power $\sim$40 W.  The trigger and data compression electronics will also consume substantial power thus a total power consumption of payload of $\lesssim$80 W is anticipated. Other readout ideas e.g. ICCD (with potential rate and weight limitation) could also be considered. 
% We note that with some weight and complexity penalty, it could be possible to apply one fiber to two measurement locations with some distance apart. Ambiguity is negligible because of the low occupancy. Power consumption of the ASICs can be reduced by a factor of 2.

\section{Mission Concept}

The mission of {\it PANGU} is a high resolution gamma-ray detector implemented with a fully active tracker and spectrometer. Being much lighter than the Fermi Gamma-ray Space Telescope (by a factor of $\sim$20 in terms of total weight), {\it PANGU} with an estimated maximal weight of satellite 250 kg, can be launched into an equatorial (inclination $<$ 5$\degree$) low Earth orbit ($\sim$550 km altitude), which offers good space environmental conditions for a gamma-ray experiment, or to L2 for stable thermal and cosmic ray environment. The lifetime of the mission should be at 2-3 years but the instrument itself can last more than 10 years. The satellite should be able to work in both the survey mode and pointing mode. Telemetry requirement is to be evaluated but is not expected to be a problem if efficient onboard data compression is used since typical photon events have relatively low occupancy (several hits per layer). Given the large field of view of the instrument, simple scanning strategy will achieve full sky coverage at $\lesssim$10 MeV to $\gtrsim$1 GeV at each revolution (i.e. every 90 minutes), similar to Fermi. 
%% The mission operations, that will take place from ESA European Space Operation Centre (or NSSC??), including the satellite control, the instrument TCs uploads and the telemetry downloading every orbit (every 90 hours, if this correct?). The content of the onboard memory will be downloaded at a rate of 100 Mbit/s. The total power budget of the {\it PANGU} mission will amount to 50 W and will be provided by (one 10 m GaAs Solar Array Wing equipped with a Solar Array Drive Mechanism.)?

%% satellite platform?

%% The Science Data Center will be in charge of the instrument health monitoring, the quick look analysis and the data processing, archiving and distribution. After a year period allocated to the {\it PANGU} team to fix the analysis software and exploit the first results, all data will become public as soon as possible.

\section{Conclusions}

We have described the {\it PANGU} project proposed as a candidate mission for the joint ESA-CAS small space mission. The {\it PANGU} instrument concept will provide unique observations in the $\lesssim$10 MeV to $\gtrsim$1 GeV energy range with unprecedented spatial resolution, almost one order of magnitude improvement in the point spread function over two order of magnitude in energy range. The {\it PANGU} instrument will open up a new window in sub-GeV gamma-ray astrophysics with unique capability to measure polarization. It enables a clear separation between diffuse gamma-ray emission from point sources, thus significantly improve the systematics control from background contamination to the underlining physics. For the first time, spaceborne gamma-ray detector achieves comparable angular resolution to ground-based Cherenkov telescope at $\sim$GeV. It will address a wide range of important questions in both astronomy and fundamental physics. The few instrumental challenges are mainly engineering and readily tractable. 

As a demonstration of the validity of the adopted technology, we have reported preliminary results on  numerical simulations on the instrument performance. We will proceed with prototype detectors to determine the optimum electron energy determination algorithms, gamma-ray direction and energy, and the energy dependent polarization modulation factor. We envision {\it PANGU} as the first step opening the next generation of future space gamma-ray mission. One option would be to combine the {\it PANGU} tacker with a high energy cosmic ray experiment, such as HERD\cite{HERD}.

{\it Very interesting and well-defined development work on silicon and scintillating fiber tracker, large density readout electronics, magnet technology and system integration might draw a group of people from different communities with common interest into a future collaboration not only for PANGU mission, but also, more importantly, for developing mission concepts and payload technologies for future missions. }  

%%%%%%%%%%%%%%%%%%%%%%%%%%%%%%%%%%%%%%%%%%%%%%%%%%%%%%%%%%%%%
\acknowledgments     %>>>> equivalent to \section*{ACKNOWLEDGMENTS}       
Support for the work of M.S. was provided by NASA through Einstein Postdoctoral Fellowship grant number PF2-130102 awarded by the Chandra X-ray Center, which is operated by the Smithsonian Astrophysical Observatory for NASA under contract NAS8-03060.

%%%%%%%%%%%%%%%%%%%%%%%%%%%%%%%%%%%%%%%%%%%%%%%%%%%%%%%%%%%%%
%%%%% References %%%%%

\bibliography{spie}
\bibliographystyle{spiebib}

\end{document}